\documentclass[12pt]{article}
\usepackage{amsmath}
\usepackage{amssymb}
\usepackage[titletoc]{appendix}
\begin{document}
\vskip 2cm
\begin{center}
{\sf {\Large   $\mathcal{N }= 4$ Supersymmetric Quantum Mechanical Model: Novel Symmetries}}

\vskip 3.0cm

{\sf \large {S. Krishna} \\
\vskip .2cm
{\it Indian Institute of Science Education and Research Mohali, \\
 Sector 81,  SAS Nagar,  Manauli - 140306 (Punjab),  India}\\


{\small {\sf {e-mail: skrishna.bhu@gmail.com;  
}}}}

\end{center}

\vskip 2cm

\noindent
{\bf Abstract:} 
We discuss a set of novel discrete symmetry transformations  of the  
$\mathcal{N} = 4$ supersymmetric quantum mechanical model of a charged particle  moving
on a sphere in the background of  Dirac  magnetic monopole.  The usual {\it five} continuous symmetries 
(and their conserved Noether charges) and {\it two}
discrete symmetries  {\it together} provide the physical realizations  of the de Rham cohomological 
operators of differential geometry. We have also exploited the supervariable approach to 
 derive the  nilpotent $\mathcal{N} = 4$ SUSY transformations and  provided the geometrical 
interpretation in the language of translational generators along the Grassmannian directions
$ \theta^\alpha$ and $\bar\theta^\alpha$ onto (1, 4)-dimensional supermanifold.



\newpage
\section{Introduction}

It is a well-known  fact that
{\it three} out of {\it four} fundamental interactions  of nature   are theoretically described by
the gauge theories. 
 These theories are 
characterized by the {\it local} gauge symmetries at the {\it classical} level which are generated by the first-class constraints 
in the language of Dirac's prescription for the classification scheme [1,2].  Some of the gauge theories  provide 
the physical examples of the Hodge theory within the framework of Becchi-Rouet-Stora-Tyutin (BRST) formalism
where the local gauge symmetries of the {\it classical} theory are traded with the nilpotent (anti-) BRST and (anti-)co-BRST symmetries 
at the {\it quantum} level. In a recent paper (see, e.g. [3] for more details), we have shown that any 
 Abelian $p$-form ($p = 1, 2, 3,...$) gauge theory is a tractable model for the Hodge theory in 
$D = 2p$ dimensions of spacetime within the framework of BRST formalism.

In an earlier work [4],  the 2D {\it free} (non-)Abelian 1-form
gauge theories (without any interaction with matter fields)  have been studied and they have been shown to be a  
new class of topological 
field theories TFTs which capture some salient feature     
of Witten type and a few key feature of Schwarz-type TFTs (see, e.g. [4-7] for more details).
Furthermore, it has been shown that the 2D Abelian $U(1)$ gauge theory, interacting with Dirac fields
[8,9], is a perfect model for the {\it Hodge theory} within framework of BRST formalism.
In such kind of studies, we have shown that the 2D modified version of Proca theory  
and 6D Abelian 3-form gauge theory [10,11]  are also the perfect examples of the Hodge theory. 
In a very recent set of papers (see, e.g. [12-14] for more details), 
a collection of $\mathcal{N} = 2$ supersymmetric quantum mechanical models    have  {\it also} been shown to represent  the  models for the Hodge theory.

In our earlier  works [15-17], we have applied  supervariable approach for the derivation of supersymmetric (SUSY) transformations
for the $\mathcal{N} = 2$ SUSY quantum mechanical models (QMMs) which is a {\it novel} approach in the context of SUSY theories. 
We have established 
that the $\mathcal{N} = 2$ SUSY QMMs also provide a set of tractable physical examples of the  Hodge theory 
because their continuous symmetries
(and  conserved Noether charges) provide the physical realizations of the de Rham cohomological operators\footnote{On a 
compact manifold without a boundary, mathematically, there exits three  differential operators ($d, \delta, \Delta$)
which are called cohomological operators of differential geometry. These three de Rham cohomological operators 
obey the following algebra: $d^2 =\delta^2 =0, \Delta= (d + \delta)^2 = \{d, \, \delta \}, [\Delta, d]= [\Delta,\,\delta] =0$ 
where $(\delta)d$ are the (co-)exterior derivatives 
and $\Delta$ is the Laplacian operator.
The  exterior and co-exterior derivatives together  satisfy an interesting relationship: $\, d = \pm *\delta*$ 
  where $*$ is the Hodge duality operation. }  
of differential geometry [18-22] and the discrete symmetry of the theory turns out to be  the analogue of Hodge duality operation.
It has been demonstrated that
 the algebra of the continuous symmetries (and their conserved Noether charges) for the $\mathcal{N} = 2$ SUSY QMMs is exactly similar to 
 the Hodge algebra obeyed by the de Rham cohomological operators of differential geometry.

In a very recent set of our works  [23,24], 
we have shown that the free  version as well as  interacting $\mathcal{N} = 2$
SUSY QM model of a charged particle 
moving on a sphere  (in the background Dirac magnetic monoploe)  provide 
a set of physical examples of the Hodge theory. This model has  also been studied by others [25,26] in a different 
context. We have also shown in our works [23,24] that one can provide the geometrical meaning to 
$\mathcal{N} = 2$ SUSY transformations in the language of translational generators 
($\partial_{\theta}, \partial_{\bar\theta}$) along the Grassmannian   
directions ($\theta$, $\bar\theta$) of the (1, 2)-dimensional super-submanifolds  on which the ordinary 
$\mathcal{N} = 2$ SUSY quantum theory [25] is generalized. 
We have also shown that our $\mathcal{N} = 2$ SUSY QMMs
are the physical examples of the Hodge theory.

The main motivations of our present investigation are as follows. First, we shall prove that the $\mathcal{N} = 4$ SUSY quantum mechanical model
of a charged particle on a sphere, in the background of Dirac magnetic monopole is  a {\it perfect } model for the the Hodge theory.
Second, we shall provide the physical realizations of the de Rham cohomological operators of differential 
geometry in the language continuous symmetries (and their conserved Noether  charges) and a set of {\it novel} discrete symmetries  of our present theory.
Finally, we shall apply supervariable approach for the derivation of SUSY transformations by exploiting chiral and anti-chiral SUSY invariant restrictions and show the invariance of the Lagrangian (and its geometrical interpretation in the language 
translational generators ($\partial_{\theta^\alpha}, \partial_{\bar\theta^\alpha}$) along the Grassmannian directions ($\theta^\alpha$, $\bar\theta^\alpha$) of the (anti-)chiral super-submanifolds, in our present SUSY  theory).

The contents of our present endeavor  are as follows. In section 2, we discuss the continuous symmetries (and their conserved Noether charges)
of the Lagrangian for the $\mathcal{N} =4$ SUSY QM model of a charged particle moving on a sphere in the background Dirac magnetic monoploe. Our section 3 is devoted to the discussion of  a set of novel  discrete symmetry  transformations.
In  section 4, we lay emphasis on the algebraic structure of the  $\mathcal{N} = 4$ SUSY symmetries and corresponding conserved charges.
Our section 5 is devoted to the derivation for the $\mathcal{N} = 4$ SUSY transformations ($s_\alpha$, $\bar s_\alpha$) 
 by exploiting SUSY invariant restrictions
within the framework of supervariable approach. 
 Finally, we make some concluding remarks in our section 6.

In our Appendix A, we discuss the explicit computations of $\mathcal{N} = 4 $ SUSY QM algebra
for the  generators ($Q_\alpha, \bar Q_\alpha$) and corresponding Hamiltonian $H$ by exploiting  the symmetry transformations
in our present SUSY theory. We provide the key difference between the (anti-)BRST symmetries  and the $\mathcal{N} = 4$ SUSY transformations  in
our Appendix B.

{\it Notations and Convention:} We adopt the following notations and convention  of  the  Grassmanian variables $\theta^\alpha$ and
$\bar\theta^\alpha$  such as:
$\{\theta^\alpha, \theta^\beta \} = 0 = ({\theta^\alpha})^2  \Rightarrow (\theta^1)^2 = (\theta^2)^2 = 0 $, 
$\{\bar\theta^\alpha, \bar\theta^\beta \} =0 =({\bar\theta^\alpha})^2 
 \Rightarrow ({\bar\theta}^1)^2 = ({\bar\theta}^2)^2 = 0, \{\theta^\alpha,\, \bar\theta^\beta \} = 0$. Similarly,
 $\{\partial_{\theta^\alpha}, \partial_{\theta^\beta} \} = 0 = ({\partial_{\theta^\alpha}})^2  \Rightarrow ({\partial_{\theta^1}})^2 
= (\partial_{\theta^2})^2 = 0 $, 
$\{\partial_{\bar\theta^\alpha}, \partial_{\bar\theta^\beta} \} =0 =({\partial_{\bar\theta^\alpha}})^2 
 \Rightarrow (\partial_{{\bar\theta}^1})^2 = (\partial_{{\bar\theta}^2})^2 = 0, \{\partial_{\theta^\alpha},\, \partial_{\bar\theta^\beta} \} = 0$
where $\partial_{\theta^\alpha} = \frac{\partial}{\partial\theta^\alpha}, \, \partial_{\bar\theta^\alpha} = \frac{\partial}{\partial {\bar\theta}^\alpha}$.

\section{Preliminaries: $\mathcal{N} =4$ SUSY Symmetries  }

Let us begin with the   Lagrangian for the $\mathcal{N} = 4$ SUSY quantum 
mechanical model of the motion of an electron   on  a  sphere in the background of Dirac
magnetic monopole based on the $CP^{(1)}$-model approach  (see, e.g.  [26] for more details)       
\begin{eqnarray}
L &=& 2\, (D_t {\bar z}) \cdot (D_t z) + \frac{i}{2}\, \left[ \bar\psi_\alpha \cdot (D_t\psi_\alpha)
- (D_t {\bar\psi}_\alpha)\cdot \psi_\alpha \right] \nonumber\\
&-& \frac{1}{4}\, \left[(\epsilon_{\alpha\beta}\,\bar\psi_\alpha\cdot\psi_\beta)^2 + 
(\bar\psi_\alpha\cdot\psi_\alpha)^2 \right] - 2\, g\, a, 
\end{eqnarray}
where the covariant derivatives  are  defined as:  $\ D_t \bar z= (\partial_t 
+ i\,a )\,\bar z, \, D_t z = (\partial_t - i\,a )\,z,\,D_t \bar\psi_\alpha = (\partial_t + i\,a )
\,\bar\psi_\alpha, \, D_t\psi_\alpha=  (\partial_t - i\,a )\,\psi_\alpha$. Here $a$ is the ``gauge" variable and $t$ is the evolution parameter  (with $\partial_t = {\partial}/{\partial t}$)
of our present  SUSY theory.
The dynamical  variables $z$ and $\bar z$ are bosonic 
in nature and the variables  $\psi_\alpha$ and  $\bar \psi_\alpha$ are the fermionic in nature (i.e. $\psi^2_\alpha = \bar\psi^2_\alpha = 0, 
\psi_\alpha\cdot\bar\psi_\beta + \bar\psi_\beta\cdot\psi_\alpha = 0$) at the classical level (with $\alpha, \beta,... = 1, 2$). 
The parameter $g$ 
stands for the charge on the magnetic monopole (with mass $m =1$) and charge of the electron is taken to be $e = -1$. 
We adopt the following conventions of the dot product 
 between two bosonic  variables ($\bar z_i$, $z_j$) 
and two fermionic 
variables ($\bar\psi_{\alpha i}$, $\psi_{\alpha j}$)  ($i,j = 1, 2; \alpha, \beta,... = 1, 2 $) are as follows:
\begin{eqnarray}
&& z_i = \begin{pmatrix}  z_1 \\ z_2 \end{pmatrix}, \qquad \bar z_j = \begin{pmatrix} \bar z_1 & \bar z_2 \end{pmatrix}
\Longrightarrow \bar z\cdot z = {\bar z}_1 z_1 + {\bar z}_2 z_2, \nonumber\\
&&
\bar\psi_{\alpha i} \psi_{\alpha j}\quad\;\, \Longrightarrow\;\; \bar\psi_\alpha \cdot \psi_\alpha = \bar\psi_1 \cdot\psi_1 
+ \bar\psi_2 \cdot \psi_2.
\end{eqnarray}
Here the fermionic variables $\bar\psi_{\alpha i}$ and $\psi_{\alpha j}$    satisfy the following properties: 
$(\bar\psi_\alpha\cdot \psi_\alpha)^2 = (\bar\psi_1\cdot \psi_1 + \bar\psi_2\cdot\psi_2)^2 
\equiv 2(\bar\psi_1 \cdot\psi_1)(\bar\psi_2 \cdot\psi_2)$
and $(\epsilon_{\alpha\beta}\bar\psi_\alpha\cdot \psi_\beta)^2 = (\bar\psi_1 \cdot\psi_2 - \bar\psi_2 \cdot \psi_1)^2 \equiv  - 2(\bar\psi_1 \cdot \psi_2)(\bar\psi_2 \cdot\psi_1)$ (with $(\bar\psi_1 \cdot\psi_1)^2
= (\bar\psi_2 \cdot\psi_2)^2 = 0, (\bar\psi_1\cdot\psi_2)^2 = (\bar\psi_2 \cdot\psi_1)^2 =0$) at the classical level.

The infinitesimal, continuous and nilpotent ($s_\alpha^2 = {\bar s}_\alpha^2 = 0$) 
$\mathcal{N} = 4$ SUSY transformations ($s_\alpha\,, \, {\bar s}_\alpha$)  of the Lagrangian (1) are 
\begin{eqnarray*}
&& s_\alpha\, z = \frac{\psi_\alpha}{\sqrt{2}}, \qquad s_\alpha\, \psi_\beta =0, \qquad s_\alpha\, 
\bar\psi_\beta = \frac{2\,i\, \nabla_{\alpha\beta}\, \bar z} {\sqrt{2}}, \qquad s_\alpha\,\bar z = 0,
\nonumber\\ && s_\alpha\,(D_t\,z) = \frac{D_t\, \psi_\alpha}{\sqrt{2}},\qquad \qquad s_\alpha\,(D_t\,\bar z) = 0, 
\qquad\qquad s_\alpha \,a = 0,\nonumber\\
\end{eqnarray*} 
\begin{eqnarray}
&& {\bar s}_\alpha\, \bar z =  \frac{\bar\psi_\alpha}{\sqrt{2}}, \qquad {\bar s}_\alpha\, \bar\psi_\beta = 0,
 \qquad {\bar s}_\alpha\, \psi_\beta 
= \frac{2 \,i\, \nabla_{\alpha\beta}\, z}{\sqrt{2}}, 
\qquad {\bar s}_\alpha \, z = 0, \nonumber\\ 
&&  {\bar s}_\alpha\,(D_t\, \bar z) = \frac{ D_t\, \bar\psi_\alpha}{\sqrt{2}}, 
\qquad\qquad {\bar s}_\alpha\,(D_t\, z) = 0,\qquad \qquad {\bar s}_\alpha \,a = 0,
\end{eqnarray}
where,
\begin{eqnarray}
\nabla_{\alpha\beta} \bar z =  \delta_{\alpha\beta}\, D_t \bar z - \frac{i}{2}\, (\bar \psi_\beta\cdot \psi_\alpha -
\delta_{\alpha\beta}\, \bar\psi_\gamma \cdot\psi_\gamma)\,\bar z, \nonumber\\
\nabla_{\alpha\beta}  z =  \delta_{\alpha\beta}\, D_t  z + \frac{i}{2}\, (\bar \psi_\alpha\cdot \psi_\beta -
\delta_{\alpha\beta}\, \bar\psi_\gamma \cdot\psi_\gamma)\, z.
\end{eqnarray}
The gauge variable $a$ is defined as
\begin{eqnarray}
a = - \frac{i}{2}\, (\bar z \cdot\,\dot z - \dot{\bar z} \cdot z) - \frac{1}{2}\, (\bar\psi_\alpha \cdot\psi_\alpha),
\end{eqnarray}
which is invariant 
under the $\mathcal{N} = 4$ SUSY transformations $s_\alpha$ and ${\bar s}_\alpha$ (i.e. $s_\alpha\, a = {\bar s}_\alpha\, a = 0$)  due to the following constraints (see, e.g. [26] for more details), namely;  
\begin{eqnarray}
\bar z \cdot z - 1 = 0, \qquad \qquad \bar z \cdot \psi_\alpha = 0, \qquad\qquad   \bar\psi_\alpha \cdot z = 0.
\end{eqnarray}
The Lagrangian (1) transforms to the total time  derivatives  as   
\begin{eqnarray}
s_\alpha\, L = \frac{d}{dt}\,\left[\frac{(D_t\,\bar z)\cdot\psi_\alpha}{\sqrt{2}} \right], \qquad\qquad {\bar s}_\alpha\, L 
= \frac{d}{dt}\,\,\left[\frac{ \bar\psi_\alpha \cdot (D_t\, z)}{\sqrt{2}}\right],   
\end{eqnarray}
which demonstrates the invariance of the action integral  $S = \int dt\, L$.

We obtain the bosonic symmetry ($s^{\omega}_{\alpha\beta}$)  for the $\mathcal{N} = 4$ SUSY 
transformations $s_\alpha$ and $\bar s_\alpha$ which is nothing but 
the anticommutator of $s_\alpha$ and ${\bar s}_\alpha$ (i.e. $s^{\omega}_{\alpha\beta} = \{s_\alpha, \, {\bar s}_\beta\}$).
 The bosonic symmetry of the dynamical variables 
$z, \bar z,\psi_\alpha$ and $\bar\psi_\alpha$  are
\begin{eqnarray}
&& s^{\omega}_{\alpha\beta} \, z =  \nabla_{\alpha\beta}  z \equiv  \delta_{\alpha\beta}\, D_t  z 
+ \frac{i}{2}\, (\bar \psi_\alpha\cdot \psi_\beta -
\delta_{\alpha\beta}\, \bar\psi_\gamma \cdot\psi_\gamma)\, z, \nonumber\\
&& s^{\omega}_{\alpha\beta} \, \bar z =  \nabla_{\alpha\beta} \bar z \equiv  \delta_{\alpha\beta}\, D_t  \bar z 
- \frac{i}{2}\, (\bar \psi_\beta\cdot \psi_\alpha -
\delta_{\alpha\beta}\, \bar\psi_\gamma \cdot\psi_\gamma)\, \bar z, \nonumber\\
&& s^{\omega}_{\alpha\beta} \, \psi_\gamma =  \nabla_{\alpha\beta}  \psi_\gamma \equiv  \delta_{\alpha\beta}\, D_t  \psi_\gamma 
+ \frac{i}{2}\, (\bar \psi_\alpha\cdot \psi_\beta -
\delta_{\alpha\beta}\, \bar\psi_\rho \cdot\psi_\rho)\,\psi_\gamma, \nonumber\\
&& s^{\omega}_{\alpha\beta} \, \bar\psi_\gamma =  \nabla_{\alpha\beta}  \bar\psi_\gamma \equiv  \delta_{\alpha\beta}\, D_t  \bar\psi_\gamma 
- \frac{i}{2}\, (\bar \psi_\beta\cdot \psi_\alpha -
\delta_{\alpha\beta}\, \bar\psi_\rho \cdot\psi_\rho)\,\bar\psi_\gamma,
\end{eqnarray}
modulo a factor of ($i$). Under the above $\mathcal{N} = 4$ SUSY transformations (8), the starting Lagrangian $L$ transforms to a total
"time" derivative as:
\begin{eqnarray}
 s^{\omega}_{\alpha\beta} L &=& \frac{d}{dt} \left[ \delta_{\alpha\beta} (L + 2 g a)\right] \equiv 
\delta_{\alpha\beta} \,\frac{d}{dt} \, \Big[2 (D_t {\bar z}) \cdot (D_t z) 
+ \frac{i}{2}\, \{ \bar\psi_\gamma \cdot (D_t\psi_\gamma)
\nonumber\\ & -& (D_t {\bar\psi}_\gamma)\cdot \psi_\gamma \}  - \frac{1}{4}\, \left\{(\epsilon_{\gamma\rho}\,\bar\psi_\gamma\cdot\psi_\rho)^2 + 
(\bar\psi_\gamma\cdot\psi_\gamma)^2 \right\} \Big].
\end{eqnarray}
As a consequence, the corresponding action  ($S = \int dt \, L$)
 remains invariant under the above bosonic symmetry transformations  ($s^{\omega}_{\alpha\beta}$) of the $\mathcal{N} = 4$ SUSY QM model.

According to Noether's theorem, the above continuous symmetry transformations ($s_\alpha$, ${\bar s}_\alpha$, 
 $s^{\omega}_{\alpha\beta}$) lead to the derivation of the following  conserved charges
\begin{eqnarray}
 Q_\alpha &=&  \frac{\Pi_z  \cdot \psi_\alpha}{\sqrt{2}} = \frac{1}{\sqrt{2}}\,\left[2\, D_t\, \bar z 
+ \frac{i}{2}\, (\bar\psi_\gamma\cdot \psi_\gamma + 2\, g)\, \bar z + 2\,i\,a\,\bar{z}\,(1 
- \bar{z}\cdot z)\right]\cdot \psi_\alpha \nonumber\\
&\equiv & 
\frac{1}{\sqrt{2}}\,\left[ 2\, D_t\, \bar z 
+ \frac{i}{2}\, (\bar\psi_\gamma\cdot \psi_\gamma + 2\, g)\, \bar z \right]\cdot \psi_\alpha, \nonumber\\
 {\bar Q}_\alpha &=& \frac{\bar\psi_\alpha \cdot  \,\Pi_{\bar z}}{\sqrt{2}} =\frac{1}{\sqrt{2}}\,  \bar\psi_\alpha
\cdot \left[2\, D_t \, z - \frac{i}{2}\, (\bar\psi_\gamma \cdot\psi_\gamma + 2\, g)\,  z
 - 2\,i\,a\,z\,(1 - \bar{z}\cdot z)\right] \nonumber\\
& \equiv &
 \frac{1}{\sqrt{2}}\; \bar\psi_\alpha
\cdot \left[ 2\, D_t \, z - \frac{i}{2}\, (\bar\psi_\gamma \cdot\psi_\gamma + 2\, g)\,  z \right], \nonumber\\
 Q^{\omega}_{\alpha\beta} &=&   \delta_{\alpha\beta}  \, \Big[2 (D_t {\bar z}) \cdot (D_t z) 
- g\,(\bar\psi_\gamma \cdot \psi_\gamma) + \frac{1}{4}\, \left\{(\epsilon_{\gamma\rho}\,\bar\psi_\gamma\cdot\psi_\rho)^2 - 
(\bar\psi_\gamma\cdot\psi_\gamma)^2 \right\} \Big] \nonumber\\
&\equiv & \delta_{\alpha\beta}\, H,
\end{eqnarray} 
where $H$ is the Hamiltonian of our $\mathcal{N} = 4$ SUSY QM model.  The above Hamiltonian can be derived
from the Legendre  transformations as:
\begin{eqnarray}
H &=&  \Pi_z \cdot \dot z + \dot {\bar z} \cdot \Pi_{\bar z} - \Pi_{\psi_\alpha} \cdot \dot \psi_\alpha 
+ \dot {\bar \psi}_\alpha \cdot \Pi_{\bar\psi_\alpha} - L \nonumber\\
& \equiv & 2 (D_t {\bar z}) \cdot (D_t z) -g\, (\bar\psi_\alpha \cdot \psi_\alpha)
  +  \frac{1}{4}\, 
\left[(\epsilon_{\alpha\beta}\,\bar\psi_\alpha\cdot\psi_\beta)^2 -  (\bar\psi_\alpha\cdot\psi_\alpha)^2 \right].
\end{eqnarray}
 It is elementary to check that the above canonical momenta 
 $\Pi_z$, $\Pi_{\bar z}, \Pi_{\psi_\alpha}$ and $\Pi_{\bar\psi_\alpha}$
w.r.t. the dynamical variables   $z, \bar z, \psi_\alpha$ and $\bar\psi_\alpha$   of the Lagrangian (1)  turn out to be
\begin{eqnarray}
 \Pi_z &=& \frac{\partial L}{\partial \,\dot z} = 2\, D_t\, \bar z 
+ \frac{i}{2}\, (\bar\psi_\alpha\cdot \psi_\alpha + 2\, g)\, \bar z + 2\,i\,a\,\bar{z}\,(1 - \bar{z}\cdot z), 
\nonumber\\
&\equiv &   2\, D_t\, \bar z 
+ \frac{i}{2}\, (\bar\psi_\alpha\cdot \psi_\alpha + 2\, g)\, \bar z, 
\nonumber\\
\Pi_{\bar z} &=& \frac{\partial L}{\partial \,\dot{\bar z}} =  2\, D_t \, z - \frac{i}{2}\, 
(\bar\psi_\alpha \cdot\psi_\alpha + 2\, g)\,  z - 2\,i\,a\,z\,(1 - \bar{z}\cdot z), \nonumber\\
& \equiv &  2\, D_t \, z - \frac{i}{2}\, 
(\bar\psi_\alpha \cdot\psi_\alpha + 2\, g)\,  z, \nonumber\\
 \Pi_{\psi_\alpha} & =& \frac{\partial L}{\partial \, \dot{\psi_\alpha }} \equiv - \frac{i}{2}\, \bar\psi_\alpha,
\qquad\qquad\qquad\qquad \Pi_{\bar\psi_\alpha} =  \frac{\partial L} {\partial \,\dot{\bar\psi}_\alpha}  \equiv 
- \frac{i}{2}\, \psi_\alpha,
\end{eqnarray}
where we have adopted the convention of  left derivative w.r.t. the fermionic variables
$\psi_\alpha$ and $\bar\psi_\alpha$ in the computation of $\Pi_\psi$ and $\Pi_{\bar\psi}$, respectively.

The    conservation law (i.e. $\dot{Q} = \dot{\bar{Q}} = \dot{Q}^\omega_{\alpha\beta} = 0$)
can be proven by exploiting the following equations of motion (with the constraints $\bar z\cdot z = 1, \bar z \cdot \psi_\alpha=0,
\bar\psi_\alpha \cdot z= 0$) that
emerge from the Lagrangian (1) of our  $\mathcal{N} = 4$ SUSY theory, namely;
\begin{eqnarray}
&& \frac{d\Pi_{\bar{z}}}{dt} - i\,\left[2\,a\, D_t\, z
 + \frac{\dot{z}}{2}\,(\bar{\psi}_\alpha\cdot \psi_\alpha+ 2g) \right] = 0,\nonumber\\ &&
  \frac{d\Pi_z}{dt}\, + i\, \left[2\,a\, D_t\, \bar{z} 
+ \frac{\dot{\bar{z}}}{2}\,(\bar{\psi}_\alpha\cdot \psi_\alpha+ 2g) \right] = 0, \nonumber\\ 
&& D_t\,\psi_\alpha + \frac{i}{2}\,(\epsilon_{\gamma\rho}\bar{\psi}_\gamma\cdot \psi_\rho)\,
(\epsilon_{\alpha\beta}\,\psi_\beta)-i\,g\, \psi_\alpha = 0, \nonumber\\ &&  
D_t\, \bar{\psi}_\alpha 
+ \frac{i}{2}\,(\epsilon_{\gamma\rho}\bar{\psi}_\gamma\cdot \psi_\rho)\,
(\epsilon_{\alpha\beta}\,\bar\psi_\beta)+ i g\, \bar\psi_\alpha= 0,
\end{eqnarray}
where canonical conjugate momenta $\Pi_{\bar{z}}$ and $\Pi_{{z}}$ in the above equation are from (12) of the $\mathcal{N} = 4$ SUSY
quantum mechanical system.

\noindent
\section{Novel Discrete Symmetry Transformations}

It is straightforward to check  that under  
the following discrete symmetry transformations
\begin{eqnarray}
&& z \rightarrow  \mp \,\bar z, \qquad \bar z \rightarrow   \mp\,  z, \qquad
\psi_\alpha \rightarrow  \mp\, \bar \psi_\alpha, \qquad \bar\psi_\alpha \rightarrow  \pm \, \psi_\alpha, \nonumber\\ && t 
\rightarrow -\, t,\qquad a \rightarrow + a,  \qquad g \to g, 
\end{eqnarray}
the Lagrangian (1) remains invariant. 
The time-reversal (i.e. $t \rightarrow -\,t$) symmetry implies:  $z \rightarrow \mp\,\bar z \Rightarrow  
z(t) \rightarrow z(-t) = \mp\,{\bar z}^T(t),\, \bar z \rightarrow \mp\, z \Rightarrow  
\bar z(t) \rightarrow \bar z(-t) = \mp\,{ z}^T(t)$,
$\psi_\alpha \rightarrow \mp\,\bar{\psi}_\alpha \Rightarrow \psi_\alpha(t)
 \rightarrow \psi_\alpha (-\,t) = \mp\,{\bar\psi_\alpha}^T (t),\,   \bar\psi_\alpha \rightarrow \,{\psi_\alpha}
 \Rightarrow \bar\psi_\alpha(t)
 \rightarrow \bar\psi_\alpha (-\,t) = \pm {\psi_\alpha}^T (t),\, a (t) \rightarrow a(-t) =
a (t)$, where the superscript
 $T$  denotes the transpose operations on the dynamical variables.

The above set of discrete symmetry transformations are the {\it novel}
useful symmetries because they establish a set of connections between the $\mathcal{N} = 4$ SUSY symmetry transformations  $s_\alpha$ and $\bar s_\alpha$ as
\begin{eqnarray}
 {\bar s}_\alpha  = \pm\; * s_\alpha\; *,
\end{eqnarray} 
where $*$ is the discrete 
symmetry transformations. The ($\pm$) signs in the above 
equation are governed by  two successive operations on the generic variable $\Phi = z, \bar z, \psi_\alpha, \bar\psi_\alpha$
\begin{eqnarray}
&& *\; (\; *\;   \Phi) = \; \pm\; \Phi.
\end{eqnarray}
It can be explicitly checked that 
\begin{eqnarray}
&& *\; (\; *\;   \Phi_1) = \; +\; \Phi_1,  \qquad \qquad \Phi_1 = z, \, \bar z, \nonumber\\
&& *\; (\; *\;   \Phi_2) = \; -\; \Phi_2,  \qquad \qquad \Phi_2 = \; \psi_\alpha , \; \bar \psi_\alpha.
\end{eqnarray}
Thus, we obtain the following relationships among the continuous  symmetry transformations ($s_\alpha, {\bar s}_\alpha$) and the discrete 
symmetry  ($*$) for the $\mathcal{N} = 4$ SUSY quantum mechanical system are
\begin{eqnarray}
&& {\bar s}_\alpha\, \Phi_1 = +\; * s_\alpha\; *\; \Phi_1 \;\;\Rightarrow\;\; {\bar s}_\alpha = + \;*\; s_\alpha\; *, \qquad\qquad
\Phi_1 = z, \,\bar z,   \nonumber\\
&& {\bar s}_\alpha \, \Phi_2 = -\; * s_\alpha\; *\; \Phi_2 \;\;\Rightarrow\;\; {\bar s}_\alpha = - \;*\; s_\alpha\; *, 
\qquad\qquad \Phi_2 = \psi_\alpha, \bar\psi_\alpha,
\end{eqnarray}
and it can be easily checked  that their reciprocal relationships are also true,  namely;
\begin{eqnarray}
&& s_\alpha \,\Phi_1 = - \; * {\bar s}_\alpha\; *\; \Phi_1 \;\;\Rightarrow\;\; s_\alpha = - \;*\; {\bar s}_\alpha\; *, \qquad\qquad
\Phi_1 = z, \,\bar z,   \nonumber\\
&& s_\alpha \, \Phi_2 = + \; * {\bar s}_\alpha\; *\; \Phi_2 \;\;\Rightarrow\;\; s_\alpha = + \;*\; {\bar s}_\alpha\; *. 
\qquad\qquad \Phi_2 = \psi_\alpha, \bar\psi_\alpha.
\end{eqnarray} 
The above relationships (15), (18) and (19) are the analogues of the relationship $\delta = \pm * d * $ of differential geometry
 where $d = dt\, \partial_t$ ($d^2 = 0$) is the exterior derivative, $\delta$ (with $\delta^2 = 0$) is the co-exterior derivative
and ($* $) is the Hodge duality operation on a given compact manifold. In our $\mathcal{N} = 4$ SUSY QM model, 
the discrete symmetry ($*$)  transformation is the analogue of the the Hodge duality 
operation ($*$).

In addition to the discrete symmetry transformations (14), the Lagrangian (1) remains invariant under the following
discrete symmetry transformations 
\begin{eqnarray}
&& z \rightarrow  \pm  i \,\bar z, \,\qquad \bar z \rightarrow   \mp\, i\,  z, \,\qquad
\psi_\alpha \rightarrow  \pm \, i\, \bar \psi_\alpha, \qquad
 \bar\psi_\alpha \rightarrow  \pm\, i \, \psi_\alpha, \nonumber\\ && t 
\rightarrow -\, t, \qquad a \rightarrow + \,a, \qquad g \to g,
\end{eqnarray}
which  obey all the conditions that have been satisfied by (14). 
Thus, these discrete symmetries are also {\it useful}, in our present theory.

Furthermore, under another  discrete symmetry transformations
\begin{eqnarray}
&& z \rightarrow  \pm   \,\bar z, \,\qquad \bar z \rightarrow   \pm\, \,  z, \,\qquad
\psi_\alpha \rightarrow  \pm \, \, \bar \psi_\alpha, \qquad
 \bar\psi_\alpha \rightarrow  \pm\,  \, \psi_\alpha, \nonumber\\ && t 
\rightarrow +\, t, \qquad a \rightarrow - \,a, \qquad g \to - g, 
\end{eqnarray}
the Lagrangian (1) remains unchanged. But, these symmetries are  
{\it not} acceptable to us because they do not comply with the strictures laid down by
the duality invariant theories [27].

In the above discrete symmetry (21), 
it can be checked that 
\begin{eqnarray}
 * \,(*\, z) = z, \qquad * \,(*\, \bar z) = \bar z, \qquad * \,(*\, \psi_\alpha) = \psi_\alpha, \qquad 
* \,(*\, \bar \psi_\alpha) =  \bar \psi_\alpha.
\end{eqnarray}
In view of the above equation  (22), we can verify that the following is true:
\begin{eqnarray}
 s_2 \; \Phi = + \, *\,s_1\, *\, \Phi, \qquad \qquad \Phi = z, \bar z, \psi_\alpha, \bar\psi_\alpha.
\end{eqnarray}
However, we note that the reciprocal relation
\begin{eqnarray}
 s_1 \; \Phi = - \, *\,s_2\, *\, \Phi, \qquad \qquad \Phi = z, \bar z, \psi_\alpha, \bar\psi_\alpha
.
\end{eqnarray}
is {\it not} satisfied at all by the above discrete symmetries.  Thus, the discrete symmetry transformations (21) of the Lagrangian (1)
are  {\it not} acceptable because they do not satisfy all the conditions (e.g. reciprocal relationship (24)) 
laid down by the duality invariant theories [27].

The conserved charges ($Q_\alpha,\, \bar Q_\alpha,\, Q^{\omega}_{\alpha\beta} \equiv \delta_{\alpha\beta} H$) 
under the discrete
 symmetry transformations transform  as:
\begin{eqnarray}
&& * \, Q_\alpha = -\, \bar Q_\alpha,\; \quad\qquad\qquad *\, \bar Q_\alpha = Q_\alpha,\qquad\qquad \qquad * \, H = H,\nonumber\\
&& *(* \, Q_\alpha) = -\, \bar Q_\alpha, \qquad\qquad *(*\, \bar Q_\alpha) = Q_\alpha,\qquad \qquad *(* \, H) = H,
\end{eqnarray}
As a consequence, under the discrete symmetries  the conserved charges $Q_\alpha$ and $\bar Q_\alpha$ transform as: $Q_\alpha\rightarrow - \bar Q_\alpha, \bar Q_\alpha
\rightarrow + Q_\alpha$ which is like the dulaity transformations  in the electrodynmics 
where we have: $B\rightarrow -E, E\rightarrow  +B$ for the electric and magnetic fields present in source free Maxwell's equations.
Furthermore,   these charges ($Q_\alpha, \bar Q_\alpha$) and the corresponding  Hamiltonian ($H$) remain invariant under two successive   ($*$) operations corresponding to the discrete symmetry transformations (14) and (20).
The above charges $Q_\alpha$ and $\bar Q_\alpha$ are the fermionic in nature and they obey the following $\mathcal{N} = 4$ SUSY QM algebra:
\begin{eqnarray}
&& \{Q_\alpha, \, Q_\beta\} \equiv Q^2_\alpha = 0 \;\,\quad \Longrightarrow  Q_1^2 =0, \quad Q_2^2 =0, \nonumber\\ &&
  \{\bar Q_\alpha, \, \bar Q_\beta\} \equiv {\bar Q}^2_\alpha =0 \;\,\quad  \Longrightarrow  
 {\bar Q}_1^2 =0, \quad {\bar Q}_2^2 =0, \nonumber\\ &&
 \{ Q_\alpha,\, \bar Q_\beta\} = \delta_{\alpha\beta}\, H,
 \qquad [H, \, Q_\alpha] = [H, \,\bar Q_\alpha] = 0.
\end{eqnarray}
It is the well-known  algebra for the  $\mathcal{N} = 4$ supersymmetric  quantum mechanical models. 
The above equation  (26) shows that the Hamiltonian ($H$) commutes with the charges $Q_\alpha$ and $\bar Q_\alpha$
(i.e. $[H, \, Q_\alpha] = [H, \,\bar Q_\alpha] = 0$). The anticommutator of these charges also give rise to the Hamiltonian 
(i.e. $\{ Q_\alpha,\, \bar Q_\beta\} = \delta_{\alpha\beta}\, H$) and they are  fermionic in nature 
(i.e. $Q_\alpha^2 =0, {\bar Q_\alpha^2 =0}$) which show the nilpotency property   of our $\mathcal{N} = 4$ SUSY QM system.

\noindent
\section{Algebraic structure: Towards Cohomological Aspects for the $\mathcal{N} = 4$ SUSY Symmetries 
}

The continuous symmetry transformations ($s_\alpha,\,{\bar s}_\alpha,\,s^{\omega}_{\alpha\beta}$)
 satisfy  the following algebraic structure:
\begin{eqnarray}
&& \{s_\alpha, \, s_\beta\} \equiv s_\alpha^2 = 0 \qquad \Longrightarrow s^2_1  =0,\quad s_2^2=0, \nonumber\\ && 
\{\bar s_\alpha,\, \bar s_\beta\} \equiv {\bar s}_\alpha^2 = 0 \qquad \Longrightarrow
{\bar s}_1^2 = 0,\quad {\bar s}_2^2 = 0, \nonumber\\
&& \{ s_\alpha, \,{\bar s}_\beta \}= s^{\omega}_{\alpha\beta},  
\qquad  \left[s^{\omega}_{\alpha\beta}, s_\gamma \right] = 0, \quad 
 [s^{\omega}_{\alpha\beta}, {\bar s}_\gamma ] = 0, \quad 
\{ s_\alpha, \,{\bar s}_\beta \}  \neq 0 
\end{eqnarray}
Here the bosonic symmetry transformations ($s^\omega_{\alpha\beta}$) is just like as a Casimir operator of our present  theory (because it commutes with 
all other SUSY transformations $s_\alpha$ and $\bar s_\alpha$).

We note that the conserved charges  
($Q_\alpha,\, \bar Q_\alpha,\, Q^{\omega}_{\alpha\beta}$) can also be  expressed in language {\it five} continuous symmetries
$\left((s_1,s_2)\Rightarrow s_\alpha,\,({\bar s}_1, {\bar s}_2) \Rightarrow {\bar s}_\alpha, s^\omega_{\alpha\beta} \right)$ as:
\begin{eqnarray}
&& s_\alpha\,Q_\beta = \,i\,\{ Q_\beta,\, Q_\alpha \} = 0, \qquad \qquad \qquad
{\bar s}_\alpha\, \bar{Q}_\beta 
= \,i\,\{ \bar{Q}_\beta,\, \bar{Q}_\alpha \} = 0,\nonumber\\
&& s_\alpha\,{\bar  Q}_\beta = i\,\{ \bar{Q}_\beta,\, Q_\alpha \} = i\delta_{\alpha\beta}\, H, \qquad\qquad {\bar s}_\alpha\, Q_\beta 
= i\,\{ Q_\beta, \,\bar{Q}_\alpha \} = i \delta_{\alpha\beta}\, H, \nonumber\\
&& s^{\omega}_{\alpha\beta}\, Q_{\gamma} = -\,i\, \left[Q_{\gamma}, \,\delta_{\alpha\beta}\,H \right] = 0, 
\qquad\qquad s^{\omega}_{\alpha\beta} \,\bar{Q}_\gamma 
= -\,i\, \left[\bar{Q}_\gamma,\,\delta_{\alpha\beta} \,H \right] = 0.
\end{eqnarray}
At the algebraic level, the  equations (26), (27) and (28) are reminiscent of the algebra  obeyed  by the
de Rham cohomological operators of differential geometry, namely;
\begin{eqnarray}
d^2 = 0, \qquad \delta^2 = 0, \qquad \Delta  = \{d, \, \delta \}, \qquad [\Delta, \, d]= 0,\qquad
[\Delta, \, \delta] = 0,
\end{eqnarray}
where $(\delta)d$ are the (co-)exterior derivatives and $\Delta$ is the Laplacian operator.
 We note that  the Laplacian operator $\Delta$  is the Casimir operator, 
because it commutes with all rest of the de Rham cohomological operators. 
Thus, ultimately, we observe that our $\mathcal{N} = 4$ SUSY quantum mechanical model provides the physical realizations 
of the de Rham cohomological operators of differential geometry in the language of symmetries and their conserved Noether charges.  
Hence,  our $\mathcal{N} = 4$ SUSY quantum mechanical model is the perfect  model for the Hodge theory.

\section{Derivation of $\mathcal{N} = 4$ SUSY Transformations: Supervariable  Approach}

\noindent
We derive the $\mathcal{N} = 4$ SUSY transformations $s_\alpha$ and $\bar s_\alpha$ within the framework of the supervariable 
approach. First, we focus on the derivation of the SUSY transformations $s_\alpha$ by exploiting the chiral supervariable approach
which is defined on the (1, 2)-dimensional  super submanifold\footnote{Here the ordinary 1D manifold characterized by $t$  has been
generalized  to (1, 2)-dimensional super-submanifold. The latter is characterized   by the superspace   variables
 $(t, \theta) \equiv (t, \theta^\alpha)$ (with $\alpha  = 1, 2$).  The Grassmannian variables $\theta^\alpha$ 
  satisfy the following properties such as: $(\theta^1)^2= (\theta^2)^2 = 0$.}. 
Thus,  the chiral supervariables\footnote{We have chosen here  
  the (anti-)chiral supervariables because the nilpotent $\mathcal{N} = 4$
SUSY transformations  do not anticommute 
(i.e. $\{s_\alpha, \bar s_\beta\} \ne 0$). This should be different  from the nilpotent
(anti-)BRST symmetry transformations  because (anti-)BRST symmetries are nilpotent as well as absolutely 
anticommuting  (see, e.g. [30-33]). Within
the framework of superfield approach to (anti-)BRST symmetries, the superfields are expanded along both
the Grassmannian directions ($\theta, \bar\theta$) (see, e.g. Appendix B).} expansions  in terms of the ordinary variables $\left( z(t), \bar{z}(t), \psi_\alpha(t), 
{\bar\psi}_\alpha(t) \right)$ are 
\begin{eqnarray}
&& z(t) \longrightarrow Z(t, \theta) = z (t) + \theta^\alpha\, f^1_\alpha(t),\nonumber\\
&& \bar{z}(t) \longrightarrow \bar{Z}(t, \theta) = \bar{z}(t) + \theta^\alpha\, f^2_\alpha(t),\nonumber\\
&& \psi_\alpha(t) \longrightarrow \Psi_\alpha (t, \theta) = \psi_\alpha (t) + i\,\theta^\beta\,b^1_{\alpha\beta}(t), \nonumber\\
&& \bar{\psi}_\alpha (t) \longrightarrow \bar{\Psi}_\alpha (t, \theta) = \bar{\psi}_\alpha (t) + i\, \theta^\beta\, b^2_{\alpha\beta}(t),
\end{eqnarray}
where the secondary variables ($ b^1_{\alpha\beta} (t), b^2_{\alpha\beta} (t)$) and ($f^1_\alpha (t), f^2_\alpha(t)$)  are bosonic and  fermionic 
 in nature, respectively.

For the derivation of  these secondary variables ($ b^1_{\alpha\beta} (t), b^2_{\alpha\beta} (t)$) and ($f^1_\alpha(t), f^2_\alpha(t)$) in terms of the basic variables,
we have to impose the  SUSY invariant restrictions (SUSYIRs). The following quantities are invariant under $s_\alpha$ such that 
\begin{equation}
s_\alpha\,(\psi_\beta) = 0,\quad s_\alpha\,(\bar{z})  = 0, \quad s_\alpha\, (z^T \cdot \psi_\beta)  = 0, 
\quad s_\alpha\, \left[2\,D_t\,{\bar{z}}\cdot z  + i\,{\bar\psi}_\beta \cdot \psi_\beta \right] = 0,
\end{equation}
where $z^T(t) \cdot \psi_\alpha(t) = z_1\,\psi_1 + z_2\,\psi_2$. The above SUSYIRs can be generalized onto (1, 2)-dimensional chiral super-submanifold. In this context, we obtain the following relationships:
\begin{eqnarray}
&& \Psi_\alpha (t, \theta) = \psi_\alpha (t) \;\Longrightarrow \; b^1_{\alpha\beta} (t) = 0,\qquad \qquad
 \bar{Z}(t, \theta) = \bar{z}(t) \; \;\Longrightarrow \; f^2_\alpha (t)  = 0,\nonumber\\
&& Z^T (t, \theta) \cdot \Psi_\alpha (t, \theta) = z^T (t) \cdot \psi_\alpha (t), \nonumber\\
&& 2\,D_t \bar{Z}(t, \theta)\cdot Z (t, \theta) + i\,\bar{\Psi}_\alpha (t, \theta) \cdot \Psi_\alpha (t, \theta) =
2\, D_t\,\bar{z}(t) \cdot z(t) + i\, \bar{\psi}_\alpha(t) \cdot \psi_\alpha (t). \;\;\;
\end{eqnarray}
The non-trivial solution of the above restrictions is  $f^1_\alpha (t) \propto \psi_\alpha (t) $. 
For the algebraic convenience, however, we  choose  $f^1_\alpha (t) =  {\psi_\alpha (t)}/{\sqrt{2}}$. 
For instance, we obtain  $b^2_{\alpha\beta} (t) = {2\,\nabla_{\alpha\beta}\,\bar z (t)}/ {\sqrt {2}}$ from last entity of equation  (32).

 Plugging in the value $ b^1_{\alpha\beta} (t) = 0,\,  f^2_\alpha (t)  = 0,\, f^1_\alpha (t) 
=  {\psi_\alpha (t)}/{\sqrt{2}}$ and $ b^2_{\alpha\beta} (t)
= {2\,\nabla_{\alpha\beta}\,\bar z (t)}/ {\sqrt {2}}$ into the chiral supervariable expansions (30), 
we obtain the following
\begin{eqnarray}
&& Z^{(1)}(t, \theta) = z (t) + \theta^\alpha\, \left(\frac{\psi_\alpha (t)}{\sqrt{2}} \right) \equiv z (t) 
+ \theta^\alpha\, \left(s_\alpha\,z (t) \right), \nonumber\\
&& \bar{Z}^{(1)}(t, \theta) = \bar{z} (t) + \theta^\alpha\, (0)  \equiv \bar{z}(t) 
+ \theta^\alpha\, \left(s_\alpha\,\bar{z} (t) \right), \nonumber\\
&& \Psi^{(1)}_\alpha(t, \theta) = \psi_\alpha(t) + \theta^\beta\,(0)  \equiv \psi_\alpha (t)
 + \theta^\beta\, \left(s_\alpha\,\psi_\beta (t) \right), \nonumber\\
&& \bar{\Psi}^{(1)}_\alpha(t, \theta) = \bar{\psi}_\alpha (t)
 + \theta^\beta\, \left( \frac{2\,i\,\nabla_{\alpha\beta}\,\bar{z}(t)}{\sqrt{2}} \right) \equiv \bar{\psi}_\alpha (t)
 + \theta^\beta\, \left(s_\alpha\,\bar{\psi}_\beta (t) \right).
\end{eqnarray} 
Here the superscript $(1)$ denotes the  expansions of the supervariables obtained after the application of the SUSYIRs.

Geometrically, the above chiral expansions of the supervariables obey  the following mapping in terms of $s_\alpha$ and
Grassmanian derivative $\partial/\partial\theta^\alpha$ such as 
\begin{eqnarray}
\frac{\partial}{\partial\theta^\alpha}\,\Omega^{(1)} (t, \theta) = s_\alpha\, \omega (t) 
\quad \Longrightarrow \quad s_\alpha\, \Longleftrightarrow \, \frac{\partial}{\partial\theta^\alpha}
\end{eqnarray}
where $\Omega^{(1)} (t, \theta)$ is the generic chiral  supervariable which   stands for $ Z^{(1)}(t, \theta),\, \bar{Z}^{(1)}(t, \theta),$ $\bar{\Psi}^{(1)}_\alpha(t, \theta),\,  \bar{\Psi}^{(1)}_\alpha(t, \theta) $ 
and $ \omega (t) = z(t),\bar z(t), \psi_\alpha(t), \bar\psi_\alpha (t)$ is the generic ordinary 
variable of our present  theory.

To derive the SUSY transformations ${\bar s}_\alpha$ by exploiting SUSY invariant restrictions (SUSYIRs)
 on the anti-chiral supervriables of the basic variables $(z(t), 
\bar{z}(t)$, $\psi_\alpha(t), \bar{\psi}_\alpha(t))$ onto  (1, 2)-dimensional anti-chiral super submanifold.
 The anti-chiral supervariable expansions of these basic variables are
\begin{eqnarray}
&& z(t) \; \longrightarrow \;  Z(t, {\bar\theta}) = z(t) +  {\bar\theta}^\alpha\, f^3_\alpha(t),\nonumber\\
&& \bar z(t) \; \longrightarrow \; \bar Z(t, {\bar\theta}) = \bar z(t) +  {\bar\theta}^\alpha\, f^4_\alpha(t),\nonumber\\
&& \psi_\alpha(t) \;\longrightarrow \; \Psi_\alpha (t, \bar\theta) = \psi_\alpha (t)  + i\,\bar \theta^\beta\, b^3_{\alpha\beta} (t), \nonumber\\
&& \bar\psi_\alpha (t)\; \longrightarrow  \; \bar\Psi_\alpha (t, \bar\theta) = \bar\psi_\alpha (t)  + i\, \bar\theta^\beta\, b^4_{\alpha\beta} (t),
\end{eqnarray} 
where ($f^3_\alpha (t), f^4_\alpha (t) $) and ($b^3_{\alpha\beta} (t), b^4_{\alpha\beta} (t)$) are the pair of fermionic  and bosonic 
secondary variables, respectively,  on the r.h.s. of anti-chiral supervariable  expansions (35). Here the anti-chiral 
super-submanifold  is parametrized by the
 superspace variables  
($t, \bar\theta^\alpha$) (where
${\bar\theta}^\alpha = {\bar\theta}^1, {\bar\theta}^2$).  
We obtain the following  SUSYIRs under ${\bar s}_\alpha$ 
\begin{equation}
{\bar s}_\alpha\,(\bar\psi_\beta) = 0,\quad {\bar s}_\alpha\,({z})  = 0, \quad {\bar s}_\alpha\, (\bar z \cdot {\bar\psi}_\beta^T)  = 0, 
\quad {\bar s}_\alpha\, \left[2\,{\bar{z}}\cdot D_t\, z  - i\,\bar{\psi}_\beta\cdot \psi_\beta\right] = 0.
\end{equation}
We demand the SUSY invariant quantities would be independent of the Grassmannian  variable $\bar \theta^\alpha$ on the (1, 2)-dimensional
 anti-chiral super submanifold.  The above secondary variables ($f^3_\alpha, f^4_\alpha$, $b^3_{\alpha\beta}, b^4_{\alpha\beta}$)
 can be obtained in terms 
of the basic variables if we impose SUSYIRs on the anti-chiral supervariables. 
 Thus, we impose the following SUSYIRs 
\begin{eqnarray}
&& Z (t, \bar\theta) = z (t), \qquad \bar\Psi_\alpha (t, \bar\theta)
= \bar\psi_\alpha(t), \qquad  \bar Z (t, \bar\theta)\cdot {\bar\Psi}_\alpha^T (t, \bar\theta) 
= \bar z (t)\cdot {\bar\psi}_\alpha^T (t), \nonumber\\
&& 2\,{\bar Z} (t, \bar\theta) \cdot D_t\,Z (t, \bar\theta) - i \, {\bar\Psi}_\alpha (t, \bar\theta) 
\cdot \Psi_\alpha (t, \bar\theta) =  2\,{\bar z}(t)\cdot  D_t\,z(t) - i\,{\bar\psi}_\alpha(t)\cdot\psi_\alpha(t),\quad
\end{eqnarray}
which imply the following  results after the substitution of the proper supervariable expansions  (35), namely; 
\begin{eqnarray}
\quad f^3_\alpha (t) = 0, \qquad b^4_{\alpha\beta} (t) = 0, \qquad f^4_\alpha (t)
 = \frac{\bar\psi_\alpha (t)}{\sqrt 2}, \qquad b^3_{\alpha\beta} (t) = \frac{2\, \nabla_{\alpha\beta}\,z (t)}{\sqrt 2}.
\end{eqnarray}
The substitution of the above  secondary variables (38) into the supervariable  expansions (35) after SUSYIRs
 lead to the following anti-chiral supervariable  expansions 
\begin{eqnarray}
&& Z^{(2)}(t, \bar\theta) = z(t) + \bar \theta^\alpha\,(0) \equiv z(t) 
+ \bar\theta^\alpha \,({\bar s}_\alpha\, z),\nonumber\\
&& {\bar Z}^{(2)}(t, \bar\theta) = \bar z(t) +  \bar\theta^\alpha\, \left(\frac{{\bar\psi}_\alpha}{\sqrt 2} \right) \equiv \bar z(t) 
+ \bar\theta^\alpha \,({\bar s}_\alpha\, \bar z),\nonumber\\
&& \Psi_\alpha^{(2)} (t, \bar\theta) = \psi_\alpha (t)  + \bar\theta^\beta\,\left(\frac{2\,i\,\nabla_{\alpha\beta}\,{ z}}{\sqrt 2} \right)
 \equiv \psi_\alpha (t) + \bar\theta^\beta\, (\bar s_\alpha\, \psi_\beta), \nonumber\\
&& \bar\Psi_\alpha^{(2)} (t, \bar\theta) = \bar\psi_\alpha (t)  + \, \bar\theta^\beta\,(0)  
\equiv \bar\psi_\alpha (t)  + \, \bar\theta^\beta\, ({\bar s}_\alpha\, \bar\psi_\beta),
\end{eqnarray} 
where the superscript $(2)$ denotes the expansions of the supervariables after the application of SUSYIRs
in (38).

The conserved charges $Q_\alpha$ and $\bar Q_\alpha$  can  be expressed in terms of the (anti-)chiral supervariable expansions  
after the application of SUSYIRs and can be expressed   in {\it two} different ways:
\begin{eqnarray}
 Q_\alpha &=&\frac{\partial}{\partial\theta^\alpha}\, \Big[2\,D_t {\bar Z}^{(1)}(t, \theta)
\cdot Z^{(1)}(t, \theta)\Big]  
 \equiv  \frac{\partial}{\partial \theta^\alpha}\, \Big[2\,D_t {\bar z}(t)
\cdot Z^{(1)}(t, \theta)\Big], \nonumber\\ 
 &=& \int d\theta^\alpha\, \Big[2\, D_t{\bar Z}^{(1)}(t, \theta)
\cdot Z^{(1)}(t, \theta)\Big] \equiv \int d\theta^\alpha\, \Big[2\,D_t {\bar z}(t)
\cdot Z^{(1)}(t, \theta)\Big],  \nonumber\\
Q_\alpha &=& \frac{\partial}{\partial \theta^\alpha}\, \Big[- i\,{\bar \Psi}_\beta^{(1)}(t, \theta)
\cdot \Psi_\beta^{(1)}(t, \theta)\Big]  
 \equiv \frac{\partial}{\partial \theta^\alpha}\, \Big[- i\,{\bar \Psi}_\beta^{(1)}(t, \theta)
\cdot \psi_\beta(t)\Big], \nonumber\\ 
 &=& \int d\theta^\alpha\, \Big[- i\,{\bar \Psi}_\beta^{(1)}(t, \theta)
\cdot \Psi^{(1)}_\beta(t, \theta)\Big]  
\equiv  \int d\theta^\alpha\, \Big[- i\,{\bar \Psi}^{(1)}_\beta(t, \theta)
\cdot \psi_\beta(t)\Big],  
\nonumber\\
\bar Q_\alpha &=&\frac{\partial}{\partial \bar\theta^\alpha}\, \Big[2\,{\bar Z}^{(2)}(t, \bar\theta)
\cdot D_t { Z}^{(2)}(t, \bar\theta)\Big] 
 \equiv  \frac{\partial}{\partial \bar\theta^\alpha}\, \Big[2\,{\bar Z}^{(2)}(t, \bar\theta)
\cdot D_t { z}(t)\Big], \nonumber\\ 
&=& \int d \bar\theta^\alpha\, \Big[2\,{\bar Z}^{(2)}(t, \bar\theta)
\cdot D_t { Z}^{(2)}(t, \bar\theta)\Big] 
 \equiv   \int d \bar\theta^\alpha\, \Big[2\,{\bar Z}^{(2)}(t, \bar\theta)
\cdot D_t { z}(t)\Big],  \nonumber\\
 \bar Q_\alpha &=&\frac{\partial}{\partial \bar\theta^\alpha}\, \Big[+ i\,{\bar \Psi}^{(2)}_\beta(t, \bar\theta)
\cdot \Psi^{(2)}_\beta(t, \bar\theta)\Big]  
 \equiv  \frac{\partial}{\partial \bar\theta^\alpha}\, \Big[+ i\,{\bar \psi}_\beta(t)
\cdot \Psi^{(2)}_\beta (t, \bar\theta)\Big], \nonumber\\ 
&=& \int d \bar\theta^\alpha\, \Big[+ i\,{\bar \Psi}^{(2)}_\beta(t, \bar\theta)
\cdot \Psi^{(2)}_\beta(t, \bar\theta)\Big] 
 \equiv  \int d \bar\theta^\alpha\, \Big[+ i\,{\bar \psi}_\beta(t)
\cdot \Psi^{(2)}_\beta(t, \bar\theta)\Big].
\end{eqnarray}
The nilpotency of
 ${\partial}/\partial\theta^\alpha$ and ${\partial}/{\partial{\bar\theta}^\alpha}$
(i.e. $\{\partial_{\theta^\alpha}, \partial_{\theta^\alpha}\} \Rightarrow \partial_{\theta^\alpha}^2 = 0, \,
\{\partial_{\bar\theta^\alpha}, \partial_{\bar\theta^\alpha}\} \Rightarrow \partial_{\bar\theta^2}^2 = 0$) implies that 
 $\partial_{\theta^\alpha}\, Q_\alpha = 0, \partial_{\bar\theta^\alpha}\, \bar Q_\alpha = 0$). 
The above  charges 
 $Q_\alpha$ and $\bar Q_\alpha$ can be  written  in terms of the  symmetry transformations ($s_\alpha, {\bar s}_\alpha$) and 
basic variables ($z, \bar z, \psi_\alpha, \bar\psi_\alpha$) in the following manner, namely;
\begin{eqnarray}
  Q_\alpha = s_\alpha \Big(2\, D_t {\bar z}\cdot z\Big) \equiv s_\alpha\, \Big( - i\, \bar\psi_\beta \cdot\psi_\beta\Big), \nonumber\\ 
 \bar Q_\alpha = {\bar s}_\alpha \Big(2\, {\bar z}\cdot D_t z\Big) \equiv {\bar s}_\alpha\, \Big( + i\, \bar\psi_\beta \cdot\psi_\beta \Big).
\end{eqnarray}
Thus, the above charges  $Q_\alpha$ and $\bar Q_\alpha$ are nilpotent  of order {\it two}, i.e. $Q_\alpha^2 = \frac{1}{2}\,\{Q_\alpha,\, Q_\alpha\} =0, {\bar Q}_\alpha^2 = \frac{1}{2}\,\{\bar Q_\alpha, \, \bar Q_\alpha\} =0 $ (because $s_\alpha^2 =0, {\bar s}_\alpha^2=0$ 
when we use the  constraints $\bar z \cdot z - 1= 0, \, \bar z\cdot\psi_\alpha = 0,\, \bar\psi_\alpha \cdot z = 0$).

It is straightforward to check that the invariance of the Lagrangian (1)   in terms of the (anti-)chiral supervariables 
obtained  after the application of SUSYIRs as given below 
\begin{eqnarray}
 L \; \Longrightarrow \; {\tilde L}^{(ac)} &= &2\, D_t {\bar Z}^{(2)} \cdot D_t { Z}^{(2)} 
 + \frac{i}{2}\, \left[\bar\Psi^{(2)}_\alpha \cdot D_t {\Psi}^{(2)}_\alpha  
- D_t {\bar\Psi}^{(2)}_\alpha \cdot {\Psi}^{(2)}_\alpha \right] - 2\,g\, a \nonumber\\
& \equiv & L + {\bar\theta}^{\alpha}\, \left[\frac{d}{dt}\Big(\frac{D_t {\bar z} \cdot \psi_\alpha}{\sqrt 2} \Big)\right], \nonumber\\
 L \; \Longrightarrow \; {\tilde L}^{(c)} &=& 2\,D_t {\bar Z}^{(1)} \cdot D_t { Z}^{(1)} 
 + \frac{i}{2}\, \left[\bar\Psi^{(1)}_\alpha \cdot D_ t {\Psi}^{(1)}_\alpha 
 - D_t {\bar\Psi}^{(1)}_\alpha \cdot {\Psi}^{(1)}_\alpha \right]- 2\,g\, a \nonumber\\ 
 &\equiv & L + {\theta^\alpha}\,\left[\frac{d}{dt}\Big(\frac{\bar\psi_\alpha\cdot {D_t z}}{\sqrt 2}  \Big) \right], 
\end{eqnarray}
where the superscripts $(c)$ and $(ac)$ denote the chiral and anti-chiral nature of the Lagrangians 
${\tilde L}^{(c)}_0$  and ${\tilde L}^{(ac)}_0$, respectively.
 For instance, we observe that
\begin{eqnarray}
\frac{\partial}{\partial\theta^\alpha}\; \Big[{\tilde L}^{(c)} \Big] =  \frac{d}{dt}\Big(\frac{D_t {\bar z} \cdot \psi_\alpha}{\sqrt 2} \Big) \, \equiv\, s_\alpha\, L, \nonumber\\
\frac{\partial}{\partial\bar\theta^\alpha}\; \Big[{\tilde L}^{(ac)} \Big] =  
\frac{d}{dt}\Big(\frac{\bar\psi_\alpha\cdot {D_t z}}{\sqrt 2}  \Big) \,\equiv \,{\bar s}_\alpha\, L.
\end{eqnarray}
Thus, the above relationships  provide the geometrical meaning for the  SUSY invariances
of the Lagrangian (1) in the language of the translational generators $\partial_{\theta^\alpha}$ and $\partial_{\bar\theta^\alpha}$
along the  Grassmannian discretions  $\theta^\alpha$ and $\bar\theta^\alpha$ onto (1, 4)-dimensional (anti-)chiral super-submanifolds, respectively, to produce the ordinary time derivatives
[cf. (43)] in ordinary 1D space thereby leading the symmetry invariance of our $\mathcal{N} = 4$ SUSY theory.

\section{Conclusions}

In our present endeavor, we have demonstrated that the $\mathcal{N} =4$ SUSY QMM of the motion of a charged particle
on a sphere  in the background of Dirac magnetic monoploe is a {\it perfect} model for the Hodge theory.
In this paper, we have shown that the physical realizations of the   de Rham cohomological operators ($d, \delta, \Delta$) of differential geometry  in the language of  continuous  
symmetries (and their conserved  Noether charges) and a set of novel discrete symmetries.  In addition, the discrete symmetries  
(14) and (20)  play the key role in establishing the relationships ($\bar s_\alpha = \pm * s_\alpha *, \, s_\alpha = \mp * \bar s_\alpha$) between
the continuous symmetry transformations ($s_\alpha, \bar s_\alpha$). These relations are exactly same as the relation ($\delta = \pm * d *$)
between the differential operators $d$ and $\delta$ where $*$ is the Hodge duality operation. Here the discrete symmetry is the analogue of Hodge 
duality operation. Thus,  we have shown that the perfect analogy between the de Rham cohomological operators of differential geometry
and the $\mathcal{N} = 4$ SUSY transformations (and their conserved Noether charges and Hamiltonian of the system) 
exists at the algebraic level, in our present
investigation.

In our  present endeavor, we have applied supervariable approach (we have already applied this  
approach for  different   $\mathcal{N} =2$ SUSY QMMs in [15-17,23,24])
for  the derivation of the SUSY transformations for the $\mathcal{N} = 4$ SUSY QMM
 of a charged particle (i.e. an electron)
 moving  on a sphere  in the background of Dirac magnetic monoploe  [26] within the framework of (anti-)chiral super-submanifolds.
 Similarly, we have established SUSY invariance of the Lagrangian  in the language translational of generators ($\partial_{\theta^\alpha}$, $\partial_{\bar\theta^\alpha}$) along the directions of the Grassmannian variables ($\theta^\alpha$, $\bar\theta^\alpha$) within the framework of
  chiral and anti-chiral supervariable  
 expansions (33) and (39), respectively, after imposing the SUSYIRs.

Our future endeavor is to find out the physical realizations of the de Rham cohomological operators of differential geometry
for different $\mathcal{N} = 4 $ and $\mathcal{N} = 8$ SUSY quantum mechanical models in the language of 
symmetries and conserved Noether charges. 
Furthermore, we shall apply this idea in our future investigations of the nonlinear superconformal symmetry of a fermion in the field
of a Dirac monopole  [28,29].
Our main goal is to apply  the  supervariable/superfield  approach to BRST formalism [30-34]  
 for the study of $\mathcal{N} = 2, 4, 8$ SUSY gauge theories (because of their relevance to the recent  developments in the superstring
theories), in our future publications [35]. 
 \\

\noindent
{\bf Acknowledgements:} 
We would like to gratefully acknowledge
financial support from DST, Government
of India, New Delhi, under  grant No. DST-15-0081. We would like to thank  BHU for the local hospitality during the visit.
 Fruitful suggestions  by Prof. C. S. Aulakh and Mr. T. Bhanja on the preparation  of  this present paper are also thankfully acknowledged. Enlightening comments by our esteemed Reviewer are thankfully
acknowledged, too.\\

\vskip .5cm

\noindent
{\bf{\large{~~~~~~~~~Appendix A: Symmetries and 
Algebraic Structure }}}\\

\vskip .5cm

\noindent 
In this Appendix A, we shall be  showing explicitly the $\mathcal{N} = 4$ SUSY quantum mechanical algebra (26)
amongst the conserved charges $(Q_\alpha, \bar {Q}_\alpha, Q^\omega_{\alpha\beta} \equiv \delta_{\alpha\beta} H)$ 
from  the symmetry principle. It can be explicitly checked that 
\[ s_\alpha\,Q_\beta = i\,\{Q_\beta, \, {Q}_\alpha \} = 0, \qquad \bar s_\alpha\,\bar{Q}_\beta 
= i\,\{\bar{Q}_\beta, \,\bar{Q}_\alpha  \} = 0, \eqno (A.1)\]
the l.h.s. of above equation (A.1) by using the expression for the generators ($Q_\alpha, \bar Q_\alpha$)
from (10) and the symmetry transformations from (3), we obtain the following: 
\[ s_\alpha\, Q_\beta = -\, \frac{1}{2}\, (\nabla_{\alpha\gamma} \bar z \cdot \psi_\gamma) (\bar z \cdot \psi_\beta), \qquad \qquad
\bar s_\alpha\,\bar{Q}_\beta = \, \frac{1}{2}\, (\bar \psi_\gamma \cdot \nabla_{\alpha\gamma}  z) (\bar \psi_\beta \cdot z),\eqno (A.2)\]
which turn out to be {\it zero} on the constrained surface defined by the constraint conditions  
$\bar{z}\cdot \psi_\beta = 0$ and $\bar{\psi}_\beta\cdot z = 0$, respectively. 
Similarly, we compute the l.h.s. of the following relationships:
\[s_\alpha\,\bar{Q}_\beta = i\,\{ \bar{Q}_\beta, Q_\alpha \} = i\,\delta_{\alpha\beta}\,H, \qquad \bar s_\alpha\,Q_\beta 
= i\,\{ Q_\beta, \bar{Q}_\alpha \} = i\,\delta_{\alpha\beta}\,H, \eqno (A.3)\]
by using 
the equations (10) and (3) in the above equation (A.3), we obtain the following:
\[\bar s_\alpha\,Q_\beta =   i\,\left[2 D_t \bar z + \frac{i}{2}\,(\bar\psi_\gamma \cdot \psi_\gamma + 2g) \bar z \right]\cdot \nabla_{\alpha\beta}  z
+ D_t \bar \psi_\alpha \cdot \psi_\beta  \]
 \[~~~~~~~~~ +\;\,  \frac{i}{4}\, (\bar \psi_\gamma \cdot \psi_\gamma + 2g)\, (\bar\psi_\alpha \cdot \psi_\beta) 
+ \frac{1}{2}(\bar \psi_\gamma \cdot \nabla_{\alpha\gamma} z)\,(\bar z \cdot \psi_\beta),\]
\[s_\alpha\,\bar{Q}_\beta = i\, \nabla_{\alpha\beta} \bar z\cdot\left[2\, i\,  D_t z 
 - \frac{i}{2}\,(\bar\psi_\gamma \cdot \psi_\gamma + 2g)  z \right ]
- \,( \bar \psi_\beta \cdot D_t \psi_\alpha)  \] 
\[~~~~ + \;\, \frac{i}{4}\, (\bar \psi_\gamma \cdot \psi_\gamma + 2g)\, (\bar\psi_\beta\cdot \psi_\alpha)
- \frac{1}{2}\,(\bar\psi_\beta \cdot z)(\nabla_{\alpha\gamma} \bar z \cdot  \psi_\gamma).
 \eqno (A.4)\]
 Substituting the constraints 
$\bar{\psi}_\beta \cdot z = 0, \bar z\cdot \psi_\beta = 0$ and using the definitions of $\nabla_{\alpha\beta} z, 
\nabla_{\alpha\beta} \bar z$, $ D_t z, D_t \bar z$
plus the equations of motion w.r.t. $\psi_\alpha$ and $\bar{\psi}_\alpha$  from (10) in the above equation, we obtain the following:
\[s_\alpha\,\bar{Q}_\beta =  2\, i\,\delta_{\alpha\beta}\, D_t \bar z \cdot D_t z 
+ (\bar\psi_\beta \cdot \psi_\alpha - \delta_{\alpha\beta}\bar\psi_\gamma\cdot \psi_\gamma)(\bar z \cdot \dot z-  i\, a\, \bar z\cdot z) \] 
\[~~~~~~~~~~~~~~~~~ ~~+\;  \frac{1}{2}\,\delta_{\alpha\beta}\,(\dot{ \bar z} \cdot z 
+ i\, a \, \bar z \cdot z)\, (\bar\psi_\gamma \cdot \psi_\gamma + 2\, g) 
+ \frac{i}{4}\, (\bar\psi_\gamma \cdot \psi_\gamma + 2g)\, (\bar\psi_\beta \cdot \psi_\alpha),\]
\[ - \;\,\frac{i}{4} \, (\bar\psi_\beta\cdot \psi_\alpha- \delta_{\alpha\beta}\, \bar\psi_\gamma \cdot \psi_\gamma) 
(\bar\psi_\lambda \cdot \psi_\lambda + 2g) (\bar z\cdot z) ~~~~~~~~\]
\[ +\;\, \frac{i}{2} \, (\epsilon_{\lambda\rho} \bar\psi_\lambda \cdot 
\psi_\rho)(\epsilon_{\alpha\gamma} \bar\psi_\beta \cdot \psi_\gamma) - i\,g\,
 (\bar\psi_\alpha \cdot \psi_\beta), ~~~~~~~~~~~\]
\[{\bar s}_\alpha\,{Q}_\beta =  2\, i\,\delta_{\alpha\beta}\, D_t \bar z \cdot D_t z 
- (\bar\psi_\alpha \cdot \psi_\beta - \delta_{\alpha\beta}\bar\psi_\gamma\cdot \psi_\gamma)
(\dot{\bar z} \cdot  z +  i\, a\, \bar z\cdot z) \] 
\[~~~~~~~~~~~~~~~~~ ~~ -\;  \frac{1}{2}\,\delta_{\alpha\beta}\,({ \bar z} \cdot \dot z 
- i\, a \, \bar z \cdot z)\, (\bar\psi_\gamma \cdot \psi_\gamma + 2\, g) 
+ \frac{i}{4}\, (\bar\psi_\gamma \cdot \psi_\gamma + 2g)\, (\bar\psi_\alpha \cdot \psi_\beta),\]
\[ - \;\,\frac{i}{4} \, (\bar\psi_\alpha\cdot \psi_\beta- \delta_{\alpha\beta}\, \bar\psi_\gamma \cdot \psi_\gamma) 
(\bar\psi_\lambda \cdot \psi_\lambda + 2g) (\bar z\cdot z) ~~~~~~~~\]
\[ +\;\, \frac{i}{2} \, (\epsilon_{\lambda\rho} \bar\psi_\lambda \cdot
 \psi_\rho)(\epsilon_{\gamma\alpha} \bar\psi_\gamma \cdot \psi_\beta) - i \,g\,
 (\bar\psi_\alpha \cdot \psi_\beta). ~~~~~~~~~~~~\eqno (A.5)\]
Furthermore, we use  the definition of $a$ and constraint $\bar z \cdot z = 1$ and  
$\frac{d}{dt}\,(\bar{z}\cdot z - 1) = 0$ in the above equation. We obtain the  following results
\[s_\alpha\,\bar{Q}_\beta = i\, \delta_{\alpha\beta}  \, \Big[2 (D_t {\bar z}) \cdot (D_t z) 
- g\,(\bar\psi_\gamma \cdot \psi_\gamma) + \frac{1}{4}\, \left\{(\epsilon_{\gamma\rho}\,\bar\psi_\gamma\cdot\psi_\rho)^2 - 
(\bar\psi_\gamma\cdot\psi_\gamma)^2 \right\} \Big] 
\equiv  i\, \delta_{\alpha\beta}\, H,~~~~~~~~ \]
\[\bar s_\alpha\,Q_\beta =   i\,\delta_{\alpha\beta}  \, \Big[2 (D_t {\bar z}) \cdot (D_t z) 
- g\,(\bar\psi_\gamma \cdot \psi_\gamma) + \frac{1}{4}\, \left\{(\epsilon_{\gamma\rho}\,\bar\psi_\gamma\cdot\psi_\rho)^2 - 
(\bar\psi_\gamma\cdot\psi_\gamma)^2 \right\} \Big] \nonumber\\
\equiv  i\,\delta_{\alpha\beta}\, H.  \eqno (A.6) \]
We note that, in the computation of $s_\alpha\, \bar Q_\beta = i \,\{\bar Q_\beta, \, Q_\alpha\} = i \delta_{\alpha\beta} \,H
\Rightarrow \{\bar Q_\alpha, \, Q_\beta\} =  \delta_{\alpha\beta} \,H $,
we have used the constraint conditions $\bar \psi_\beta \cdot z = 0, \bar z\cdot z = 1, 
\frac{d}{dt}\, (\bar z\cdot z - 1) = 0$ and 
the definitions of $a, \nabla_{\alpha\beta} \bar z$,  $D_t z, D_t \bar z$
plus equation of motion for $\psi_\alpha$ (i.e. $D_t\,\psi_\alpha + \frac{i}{2}\,(\epsilon_{\gamma\rho}\bar{\psi}_\gamma\cdot \psi_\rho)\,
(\epsilon_{\alpha\beta}\,\psi_\beta)-i\,g\, \psi_\alpha = 0$).  On the other hand, in the explicit composition of 
$\bar s_\alpha\,  Q_\beta = i\, \{ Q_\beta, \, \bar Q_\alpha\} = i \delta_{\alpha\beta} \,H \Rightarrow
\{ Q_\alpha, \, \bar Q_\beta\} =  \delta_{\alpha\beta} \,H $, we have exploited the constraint conditions 
$\bar z \cdot \psi_\beta = 0, \bar z\cdot z = 1, \frac{d}{dt}\, (\bar z\cdot z - 1) = 0$ 
and the definitions of $a, \nabla_{\alpha\beta} z,  D_t    z, D_t \bar z$
with using equation of motion for $\bar\psi$ (i.e. $D_t\, \bar{\psi}_\alpha 
+ \frac{i}{2}\,(\epsilon_{\gamma\rho}\bar{\psi}_\gamma\cdot \psi_\rho)\,
(\epsilon_{\alpha\beta}\,\bar\psi_\beta)+ i g\, \bar\psi_\alpha= 0$), respectively.
\\

\vskip .5cm

\noindent
{\bf{\large{Appendix A:   On the Choice of (Anti-)Chiral Supervariables for
            \vskip .1cm $~~~~~~~~~~~~~~$ the Description of SUSY Model }}}

\vskip .5cm

\noindent 
In this Appendix B, we would like to emphasize the key difference between the (anti-)chiral supervariables
in the context of the derivation of   nilpotent   SUSY transformations for the $\mathcal{N} = 4$ SUSY quantum 
mechanical model and the (anti-)BRST symmetry transformations for a gauge theory. In the literature, it is well-known 
that the (anti-)BRST symmetries ($s_{(a)b}$) for a given gauge  theory are nilpotent and absolutely anticommuting in nature 
whereas the $\mathcal{N} = 4$ SUSY symmetries are  nilpotent but {\it not} absolutely anticommuting in nature. Within the 
framework of BT-superfield approach [30-33] to 
BRST symmetries, a bosonic field $\sigma(x)$ for D-dimensional gauge theory, one has to generalize it onto a
(D, 2)-dimensional supermanifold along 
the Grassmannian directions ($\theta$ and $\bar\theta$) (with $\theta^2=\bar\theta^2=0,\; \theta\bar\theta+\bar\theta\theta =0$) in the following manner:
\[\Sigma(x,\theta, \bar\theta) =  \sigma(x) + \theta\,\bar R(x)+ \bar\theta\, R(x) + i \theta\bar\theta \, S(x),
\eqno (B.1) \]
where $R(x), \bar R(x)$ are the fermionic secondary fields and $S(x)$ is a bosonic secondary field and
 $\Sigma(x,\theta, \bar\theta)$ is the corresponding  superfield  which is  defined on the (D, 2)-dimensional supermanifold.
 In the above equation (B.1), the translational generators ($\partial_\theta, \partial_{\bar\theta}$) are found to correspond to the (anti-)BRST
symmetry transformations $s_{(a)b}$ which are nilpotent of order two due to ($\partial_{\theta}^2= \partial_{\bar\theta}^2 =0$)
 and they are absolutely anticommuting  because it is straightforward to check that:
\[\partial_{\bar\theta}\,\partial_{\theta} \, \Sigma(x,\theta, \bar\theta) = i S(x) 
\quad \Longleftrightarrow \quad s_b\,s_{ab} \sigma(x), \eqno (B.2)\]
\[ \partial_{\theta}\,\partial_{\bar\theta} \, \Sigma(x,\theta, \bar\theta) = -i S(x)
\quad \Longleftrightarrow \quad s_{ab}\,s_{b} \sigma(x). \eqno (B.3) \]
 It is clear from the above relationships (B.2) and (B.3), we obtain the following 
 \[(\partial_{\bar\theta}\,\partial_{\theta} + \partial_{\theta}\,\partial_{\bar\theta})
 \, \Sigma(x,\theta, \bar\theta) = 0
\quad \Longleftrightarrow \quad (s_b\,s_{ab} + s_b\,s_{ab})\, \sigma(x) =0, \eqno (B.4)\]
which shows  the  absolute anticommutativity  property  of the (anti-)BRST symmetry transformations. In our present 
$\mathcal{N} = 4$ SUSY QM model, we are compelled to avoid the relation  (B.4) so
that our nilpotent SUSY symmetries could not become absolutely anticommuting in nature.
The anticommutator of our present investigation (i.e. SUSY theory), is nothing but the bosonic symmetry (i.e. $s_\alpha {\bar s}_\beta + s_\beta {\bar s}_\alpha
= s^\omega_{\alpha\beta}$ with $\alpha,\beta= 1,2$) [cf. (8)]. 

Geometrically, $\mathcal{N} = 4$ SUSY symmetry transformations are identified with the translational generators ($\partial_{\theta^\alpha}, \partial_{\bar\theta^\alpha}$) along the Grassmannian
directions  ($\theta^\alpha$,
$\bar\theta^\alpha$) of the (anti-)chiral super-submanifolds which encapsulate only the nilpotency
property (not absolute anticommutativity property). The purpose of this Appendix B is to develop the theoretical tools and
techniques so that we could derive the complete structure of the SUSY symmetries for the  $\mathcal{N} = 4$ SUSY
QM model.

\end{document}